\documentclass[a4paper,twocolumn,11pt,unpublished]{quantumarticle}
\pdfoutput=1
\usepackage[utf8]{inputenc}
\usepackage[english]{babel}
\usepackage[T1]{fontenc}
\usepackage{amsmath}
\usepackage{hyperref}
\usepackage[numbers]{natbib}
\usepackage{amssymb}
\usepackage{placeins}

\begin{document}

\title{Random Number Generators in Advanced Optical Experiments: A Comparative
Analysis of Semiclassical, Quantum, and Hybrid Architectures}

\author{Daniil D.~Reshetnikov}
\affiliation{St. Petersburg State University, 7/9 Universitetskaya Nab., 199034 St. Petersburg, Russia}
\orcid{0009-0001-5587-3306}
\email{d.d.reshetnikov@gmail.com}
\author{Anna A.~Kretova}
\affiliation{St. Petersburg State University, 7/9 Universitetskaya Nab., 199034 St. Petersburg, Russia}
\orcid{0009-0000-9753-8955}
\author{Anastasia A.~Fominova}
\affiliation{St. Petersburg State University, 7/9 Universitetskaya Nab., 199034 St. Petersburg, Russia}
\orcid{0009-0001-8782-5878}
\author{Evgenii A.~Vashukevich}
\affiliation{St. Petersburg State University, 7/9 Universitetskaya Nab., 199034 St. Petersburg, Russia}
\orcid{0000-0002-5815-0103}
\author{Tatiana Y.~Golubeva}
\affiliation{St. Petersburg State University, 7/9 Universitetskaya Nab., 199034 St. Petersburg, Russia}
\orcid{0000-0002-1139-8531}
\author{Kirill S.~Tikhonov}
\affiliation{St. Petersburg State University, 7/9 Universitetskaya Nab., 199034 St. Petersburg, Russia}
\affiliation{P.N. Lebedev Physical Institute, Leninsky Prospekt 53, 119991 Moscow, Russia}
\affiliation{Russian Quantum Center, Bolshoy Bulvar 30, Bld. 1, Skolkovo IC, 121205 Moscow, Russia}
\orcid{0000-0003-2491-9727}
\email{tikhonov.kyril@gmail.com}

\maketitle

\begin{abstract}
  Random numbers sequences (RNSs) play a vital role in various scientific and engineering applications. They are critical to the integrity of classical and quantum cryptography, the accuracy of mathematical modeling and Monte Carlo simulations, and the core mechanics of applications in fields as diverse as gambling and statistical sampling. While the primary criteria for RNSs sources are their quality and generation rate, their integration into experimental designs is equally significant for many fundamental physical tests and applications. 
  
 This work presents a comparative analysis of optical random number generation architectures, which can be seamlessly included into various advanced classical and quantum optical experimental schemes. In particular, we evaluate the trade-off between the high generation rate of an attenuated laser (a quasi-single-photon source) and the superior statistical quality of a heralded single-photon source operating at a much lower frequency. To overcome the limitations of each individual source, we propose and examine a novel hybrid architecture that utilizes their mixed radiation, enabling the generation of high-quality RNSs at an enhanced rate. Furthermore, we demonstrate that the raw sequences generated by such a source can not only exhibit but, in some cases, even surpass the degree of randomness achieved by sequences processed through powerful randomness extractors.
\end{abstract}

\section{Introduction }
Large random number sequences (RNSs) and particularly Bernoulli sequences are essential for a wide range of scientific and engineering applications. This includes statistical sampling \cite{lohr_sampling_2021, wu_strong_2021, arute_quantum_2019, hangleiter_computational_2023}, computer simulations of the Monte Carlo type \cite{schmidt_james_2006, daley_quantum_2014, blanco-gonzalez_comparative_2024,dubovskaya_modeling_2025}, classical and quantum cryptography \cite{schneier_applied_2015, xu_secure_2020, portmann_security_2022}, genetic,  quantum and AI algorithms \cite{wainrib_topological_2013, klinshov_mean-field_2015, karthik_noise-resilient_2025, martyn_halving_2025, patel_ai-guided_2025, turkeshi_magic_2025, prabhu_enhancing_2025}, games of chance \cite{el-shehawey_semi-infinite_2002,szolnoki_emergence_2009,dai_double-dealing_2010,shevkoplyas_optimal_2014, szolnoki_competition_2016,rosicka_linear_2016},  and numerous others. 
RNSs can be generated through several methods, primarily falling into three categories: mathematical algorithms, physical random number generators (RNGs), or a  mixed approach combining both.  The key criteria for evaluating any of the methods are its generation rate and sequence quality, the latter being determined by the ability to pass rigorous statistical test suites like the NIST \cite{kelsey_nist_2012} or the diehard tests \cite{marsaglia_monkey_1993}. Another critical consideration is the fundamental ''nature'' of randomness itself. Specifically, whether a sequence is genuinely unpredictable or merely complex enough to appear random. The answer determines if the sequence can be deterministically reproduced given identical initial conditions -- a pivotal distinction between true random and pseudo-random generators.
In particular, purely mathematical methods generates only pseudorandom numbers. This is a direct consequence of their deterministic nature: the sequence is entirely defined by an initial ''seed'' value and a transformation algorithm. As a result, knowledge of the seed and the algorithm allows one to perfectly predict the entire sequence. This deterministic predictability renders such generators unsuitable for cryptography. Furthermore, their application in simulating inherently random natural processes, such as a time evolution of a quantum system, requires careful justification, as they only mimic randomness rather than generate it from a physical source. Despite these limitations, modern mathematical generators -- such as Collatz-Weyl Generators (CWG) and the Mersenne Twister (MT) \cite{dziala_collatz-weyl_2023} -- are highly advanced and efficient for particular tasks. They produce pseudorandom sequences with exceptionally long repetition periods (e.g., $2^{128}$ for CWG and $2^{19937}$ for MT)  and achieve high generation speeds (e.g., $\sim83$ Gbit/s and $\sim36$ Gbit/s, respectively) \footnote{The performance of software-based random number sources is fundamentally constrained only by processing speed, allowing for substantial performance gains through hardware upgrades.} . These sequences reliably pass stringent statistical tests, including the NIST and diehard suites. Consequently, being entirely software-based and highly portable these pseudorandom number generators are extensively used in a wide range of applications where the nature of randomness is not of primary importance.
Physical random number generators (RNGs) are broadly categorized as classical or quantum. Classical RNGs operate by measuring dynamical observables in macroscopic systems. A prominent example is the Field-Programmable Gate Array (FPGA) generator, which derives randomness from electrical noise to achieve production rates of several hundred Mbit/s \cite{phys_fpga_RNG,True_electr_RNG}. Alternatively, optical generators leverage phase or intensity fluctuations in laser light, enabling significantly higher rates \cite{TRNG_review}. For instance, systems utilizing a Raman fiber laser have achieved a generation rate of $540$ Gbit/s \cite{monet_simple_2021}. Even more impressive, a multichannel semiconductor laser network has reached rates up to $2.24$ Tbit/s \cite{Xiang:19}. However, the current state-of-the-art physical RNG, utilizing an integrated AlGaAs micro-resonator, has shattered previous records by achieving a tremendous speed of $126$ Tbit/s \cite{zhao_126_2025}. Although these classical sources produce high-quality RNSs, they retain a minor vulnerability: since they are rooted in classical physics, their behavior could, in principle, be predicted.
Unlike classical RNGs, Quantum Random Number Generators (QRNGs) operate by measuring the observables of a micro-object -- an entity governed by quantum mechanics. A key example is the measurement of vacuum fluctuations, a genuine random process. Through homodyne detection, this method can generate true random numbers at rates up to 100 Gbit/s \cite{bruynsteen_100-gbits_2023}, rivaling some of software-based sources and classical RNGs in speed. Other high-speed QRNG implementations, with rates ranging from hundreds of Mbit/s to several Gbit/s, exploit chaotic lasers \cite{uchida_fast_2008}, enhanced spontaneous emission \cite{williams_fast_2010}, and Raman scattering \cite{collins_random_2015}. Although the generation speed of QRNGs is often slower than that of classical alternatives, their defining advantage is the intrinsic, quantum nature of their randomness, which renders any prediction theoretically impossible.
The raw data generated by both classical and quantum RNGs typically contain biases and correlations. To convert this imperfect output into a certified random sequence, it is necessary to apply post-processing techniques known as randomness or entropy extractors.  In particular, this can be achieved through dedicated hashing steps, which enhance statistical randomness, mitigate noise, and format the sequence into fixed-length strings, or ''hashes'', for specific applications. Notable implementations of this principle include the Toeplitz matrix-based hashing \cite{goos_new_1995} and the stream numbering introduced by V.F. Babkin \cite{arbekov_extraction_2021}. Another noteworthy technique is the mixing extraction method \cite{nisan_extracting_1996}, which efficiently combines classical and quantum RNGs. This approach replaces bits in a classically-generated sequence with bits from a quantum RNG. This hybrid strategy offers a dual advantage: it disrupts any deterministic patterns from the classical source, thereby significantly improving randomness, while simultaneously achieving a much higher generation frequency than a purely quantum source.
A further important criterion of a physical RNG is its  physical footprint. The miniaturization of RNGs is essential for advancing integration, portability, and security, transforming them from bulky laboratory devices into core components of modern digital infrastructure. Here, the ultimate goal is for RNGs to function as a seamless and ''invisible'' element within a digital system. Additionally, for fundamental physical tests and many practical applications, the direct integration of an RNG in a ''circuit'' provides decisive advantages over external peripheral devices. This integrated approach eliminates the complexities of controlling and interfacing with separate hardware, thereby streamlining the entire experimental or operational workflow. Consequently, it minimizes potential sources of noise, latency, and error inherent in external connections. Furthermore, by reducing overall system complexity, integrated RNGs lower both acquisition and long-term operational costs.
In this study, we experimentally examine and compare various  optical RNGs schemes. We focus primarily on two distinct physical generators: a semiclassical source, comprising a laser attenuated to the quasi-single-photon level, and a quantum-mechanical heralded single-photon source based on parametric down-conversion. We expect, that the semiclassical source  offers a high generation rate, but its inherent photon number statistics compromise the randomness of its output. In contrast, the quantum heralded source provides a fundamentally robust random process, albeit at a significantly lower generation rate. To synthesize the benefits of these sources -- namely, high speed and certified randomness -- we propose and analyze a novel scheme of a hybrid generator in which photons from both sources are coherently mixed. Furthermore, to evaluate the randomness for each scheme we compare the raw output from the sources under consideration against pseudorandom sequences generated by the ''Python'' library ''Secrets'' and  sequences refined through the application of post-processing algorithms, specifically the von Neumann and V.F. Babkin entropy extractors. We also compare bit-mixing methods, evaluating physical mixing in the hybrid source against post-processing digital mixing techniques.  The practical merit of this work lies in the direct integrability of our proposed RNGs into broader quantum-optical experiments. For instance, they can serve as critical components in tests of Bell's inequalities \cite{greenberger_bells_1990,pironio_random_2010} that rely on polarization-encoded photons and elements of optical networks \cite{vokic_quantum_2021}. Moreover, the verification of output randomness provides a powerful tool for diagnosing experimental quality, enabling the assessment of optical alignment and the balancing of single-photon detectors. This is particularly valuable in experiments involving photon entanglement and the statistical analysis of weak optical fields.
The article is organized as follows. In Section \ref{sec: methods}, we provide a  detailed description of each RNG scheme under consideration. In Section \ref{sec: benchmarking}, we analyze the outcomes of the NIST tests for the random  sequences generated with its use. We also compare the results before and after extraction procedures with the von Neumann's and the V.F. Babkin's approaches, and propose enhancements to improve them. In conclusion \ref{sec: conclusion}, a concise synopsis of the work is provided, with an emphasis on its primary outcomes. To ensure completness of the manuscript, we have incorporated a brief description of the V.F. Babkin's stream numbering algorithm in the appendix \ref{sec: app: V.F. Babkin's extractor}. Appendix \ref{sec: app: extractor performance} provides an analysis and commentary on its effectiveness.
\section{Materials \& Methods: optical sources of random number sequences}\label{sec: methods}
 This section details the experimental methods used in our study. First, we describe a semiclassical RNG based on attenuated laser light to generate quasi-single-photon states. Next, we present a QRNG exploiting the intrinsic quantum nature of entangled photon pairs. We conclude with the design of a hybrid source that merges the outputs from the aforementioned semiclassical and quantum systems.

\subsection{A coherent quasi-single-photon source}
\begin{figure}[htbp]
    \centering
    \includegraphics[width=0.97\columnwidth]{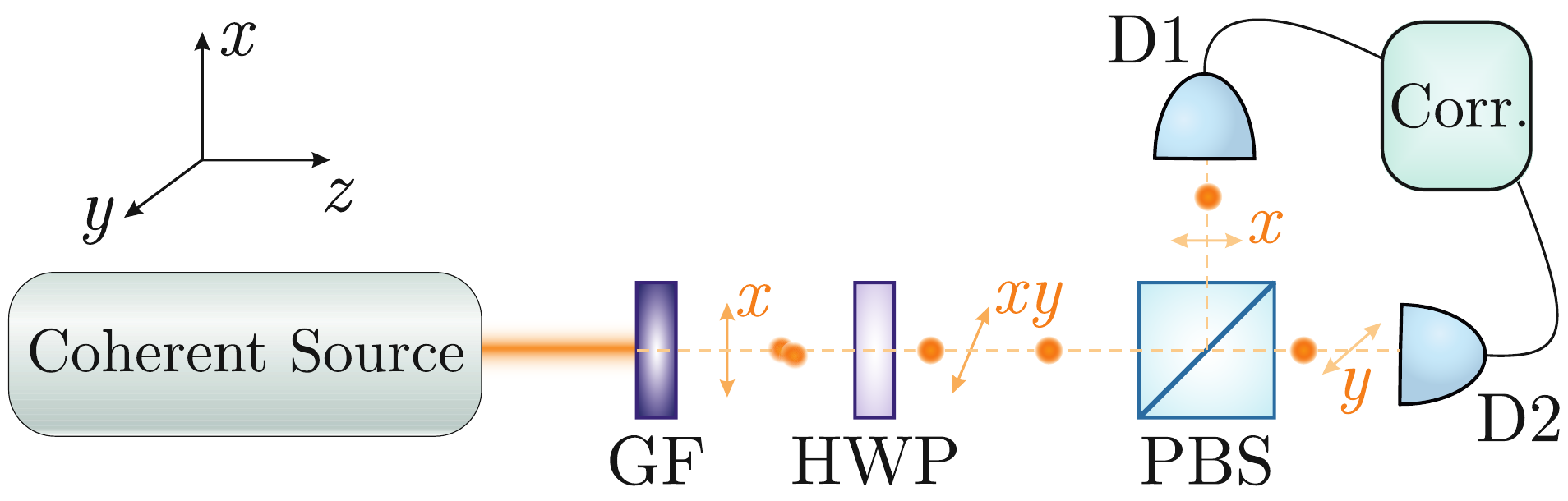}
    \caption{The coherent quasi-single-photon source of RNSs: the linearly polarized single-mode laser light (the $810$ nm semiconductor laser with a $3$ mW output power), weakened by the grey filter (GF), passes through the half-wave plate (HWP) and hits the polarizing beam splitter (PBS). After passing the beam splitter, photons are measured by detectors (D1, D2) and the corresponding signal is sent to the correlator (Corr.).}
    \label{fig1: semiclassical source}
    \end{figure}

The analysis begins by evaluating a semiclassical quasi-single-photon source for the generation of RNSs similar to \cite{argillander_tunable_2022,strydom_quantum_2024}. Its experimental scheme is illustrated in Fig.~\ref{fig1: semiclassical source}. A linearly polarized  coherent light from a single-mode semiconductor laser (FPL-810-8DL), attenuated by a grey filter (NENIR40A), passes through a half-wave plate (WPH10M) and hits a polarizing beam splitter (PBS25). After passing the beam splitter, photons are measured by detectors (COUNT® NIR) and the corresponding signal is sent to a correlator (a custom-made unit, based on the FPGA Altera Cyclone IV). The output of the first detector D1 is assigned to the binary symbol "0", while the output of the second detector D2 is assigned to the symbol "1", thereby generating a random binary sequence. 
To approximate a single-photon source, the laser output was heavily attenuated to a mean value of 0.1 photons per correlator clock cycle. At this level, the probability of a multi-photon event is only 0.5\%, ensuring that the dominant source of randomness posses quantum mechanical character (Appendix \ref{sec: app: coherent source destribution} contains the photon number distribution data). The single-photon detectors, operating at 20 MHz, is synchronized with the correlator's clock. Consequently, the maximum rate for generating random bits $R_{CS}$ is determined by the product of the detection frequency and the probability of a photon detection event $P_{DE}$, which yields a theoretical maximum of 1.9 MHz:
\begin{align}
   R_{CS}&=20\cdot10^6\cdot P_{DE}\nonumber\\&=20\cdot10^6\cdot(1-P_{0})
   \approx1.9\ \text{MHz},
\end{align}
where $P_0=0.905$ is the probability of no events occurring in a Poisson process with a mean value of 0.1.

Here, we note that, the laser attenuation was optimized to balance the random number generation rate against errors induced by multi-photon events. While higher radiation intensity increases the generation rate, it also raises the probability of multi-photon detections. Specifically, if both detectors register counts within a single clock cycle of the correlator, the source can no longer be reliably characterized as a single-photon source, compromising the integrity of the generated randomness.
%

%The quality of the generated RNS is sensitive to the alignment of the optical setup. The most salient parameters pertain to the magnitude of intensity fluctuations and the degree of ellipticity of the laser source's polarization. Fluctuations in intensity result in fluctuations in the average number of photons per pulse, thus causing the experimental setup to exit the quasi-single-photon mode. \KT{This necessitates the attenuation of light to a degree that maximally excludes two-photon states, thereby affecting the rate of generation.} The ellipticity of the polarization subsequently instigates a periodic alteration in time in the balance of signals on the first and second detectors, and consequently, the balance of ''0'' and ''1'' in the sequence.

%
Another factor impacting the quality of the generated RNSs is the polarization alignment of the optical setup. Particularly, polarization ellipticity introduces a time-dependent imbalance between the signals detected by the two photodetectors. This imperfection propagates into the output sequence, causing a periodic deviation in the balance of "0" and "1" bits, thereby compromising the sequence's randomness. The resulting bias is especially pronounced when the generated RNS is subjected to the Frequency Test from the NIST statistical test suite. The corresponding p-value $\mathcal{P}$ is calculated as follows \cite{kelsey_nist_2012}:
\begin{equation}
\mathcal{P} = \text{erfc}\ \Big((2N)^{-1/2} |D_N|\Big),    
\end{equation}
where $\text{erfc(z)}$ is the complementary error function, $D_N$ -- the discrepancy \footnote{For example, the discrepancy for the sequence $\epsilon= 1011010101$ is $S_n=1+(-1)+1+1+(-1)+1+(-1)+1+(-1)+1=2$. Here, $+1$ corresponds to ''1'' in $\epsilon$, while $-1$ corresponds to ''0''. } between the values ''0'' and ''1'' within the sequence, and $N$ -- the length of the sequence.
To ensure the test is passed (i.e. $\mathcal{P} \geq 0.01$), the balance between ''0''s and ''1''s in the final sample, defined as 
\begin{equation}
\mathcal{B}=\ (N+|D_N|)/(N-|D_N|),
\end{equation}
must meet a specific threshold since it is related to the value of $\mathcal{P}$ as
\begin{align}
\mathcal{P} = \text{erfc}\ \Big((N/2)^{1/2}\ |(\mathcal{B}-1)/(\mathcal{B}+1)|\Big).
\end{align}
This threshold is a function of the sample size $N$: for the minimum required sample of $100$ bits, the balance must be no worse than $1.7$, while for a larger sample of $10^7$ bits, a much stricter balance of $1.0016$ is required. The relationship between sample size and the permissible balance is governed by the properties of the complementary error function $\text{erfc}(z)$. This function quickly reaches a plateau, leading to the stringent balance requirement for larger sample sizes. Consequently, successfully passing the test imposes demanding constraints on the quality and precision of the optical polarization alignment.
To quantify the required optical precision, we performed a polarization state tomography on the attenuated coherent laser source. This procedure was used to identify the maximum allowable degree of polarization ellipticity and the mounting accuracy for the half-wave plate (HWP). Table I below presents the measured Stokes parameters corresponding to the minimum allowable setup accuracy for generating sequences of different lengths.
\begin{table}[h!]
\label{Table1: Stokes parameters}
\caption{The results of polarization state tomography in a setup with weak coherent laser source and varying parameters.}
\resizebox{\columnwidth}{!}{
\begin{tabular}{|r|l|l|l|l|l|}
\hline
\multicolumn{1}{|c|}{$N$} & \multicolumn{1}{c|}{$\frac{N-|D_N|}{N+|D_N|}$} & \multicolumn{1}{c|}{$S_0$} & \multicolumn{1}{c|}{$S_x$} & \multicolumn{1}{c|}{$S_y$} & \multicolumn{1}{c|}{$S_z$} \\ \hline
$10^2$                       & 1.7                                            & 1                          & 0.96399                    &  0.07258                  &  0.25584                    \\ \hline
$10^3$                      & 1.18                                           & 1                          & 0.99585                    & 0.02621                    &  0.08719                    \\ \hline
$10^4$                    & 1.05                                           & 1                          & 0.99956                    &  0.01548                   &  0.02518                   \\ \hline
$10^5$                    & 1.016                                          & 1                          & 0.99988                    &  0.01274                   &  0.00828                   \\ \hline
$10^6$                   & 1.005                                          & 1                          & 0.99991                    & 0.01261                    &  0.00254                   \\ \hline
$10^7$                  & 1.0016                                         & 1                          & 0.99992                    &  0.01258                   &   0.00082                  \\ \hline
\end{tabular}}
\end{table}
Here, the Stokes parameters $(S_x,S_y,S_z)$ are determined from the average photo-count differences between three pairs of the orthogonal polarizations: $S_x=\langle P_D\rangle-\langle P_A\rangle$  for the diagonal/anti-diagonal basis, $Sy=\langle P_R\rangle-\langle P_L\rangle$ for the right/left circular basis, and $S_z=\langle P_H\rangle -\langle P_V\rangle$ for the horizontal/vertical basis.  The total photon flux is given by $S_0=\langle P_H\rangle +\langle P_V\rangle$. $P_A=|A\rangle\langle A|$ denotes the projector onto the basis state $|A\rangle$.

The analysis of the polarization state tomography reveals that longer sample sequences impose stricter demands on the polarization of the laser radiation. Specifically, beyond a sequence length of $N=10\; 000$, this necessitates a more stable laser source and higher-quality polarizing optics. This issue is evident in the p-value distributions for the coherent source, both  in the individual test analysis (see Fig.~\ref{fig4: individual test results}a, Raw sequence) and in the aggregated test results (see Fig.~\ref{fig5: aggregated test results}a, Raw sequence). A potential solution involves implementing self-testing algorithms for long-sequence generation and active control of the apparatus parameters.
In summary, while the coherent quasi-single-photon source offers significant advantages in simplicity and speed, its output is a coherent state rather than a true single-photon state. This inherent limitation introduces discernible patterns in long raw data sequences, compromising their randomness. Consequently, for the output to pass the standard randomness tests, post-processing with randomness extractors is essential.
\subsection{A heralded (genuine) single-photon source}
\begin{figure}[htbp]
    \centering
    \includegraphics[width=0.99\columnwidth]{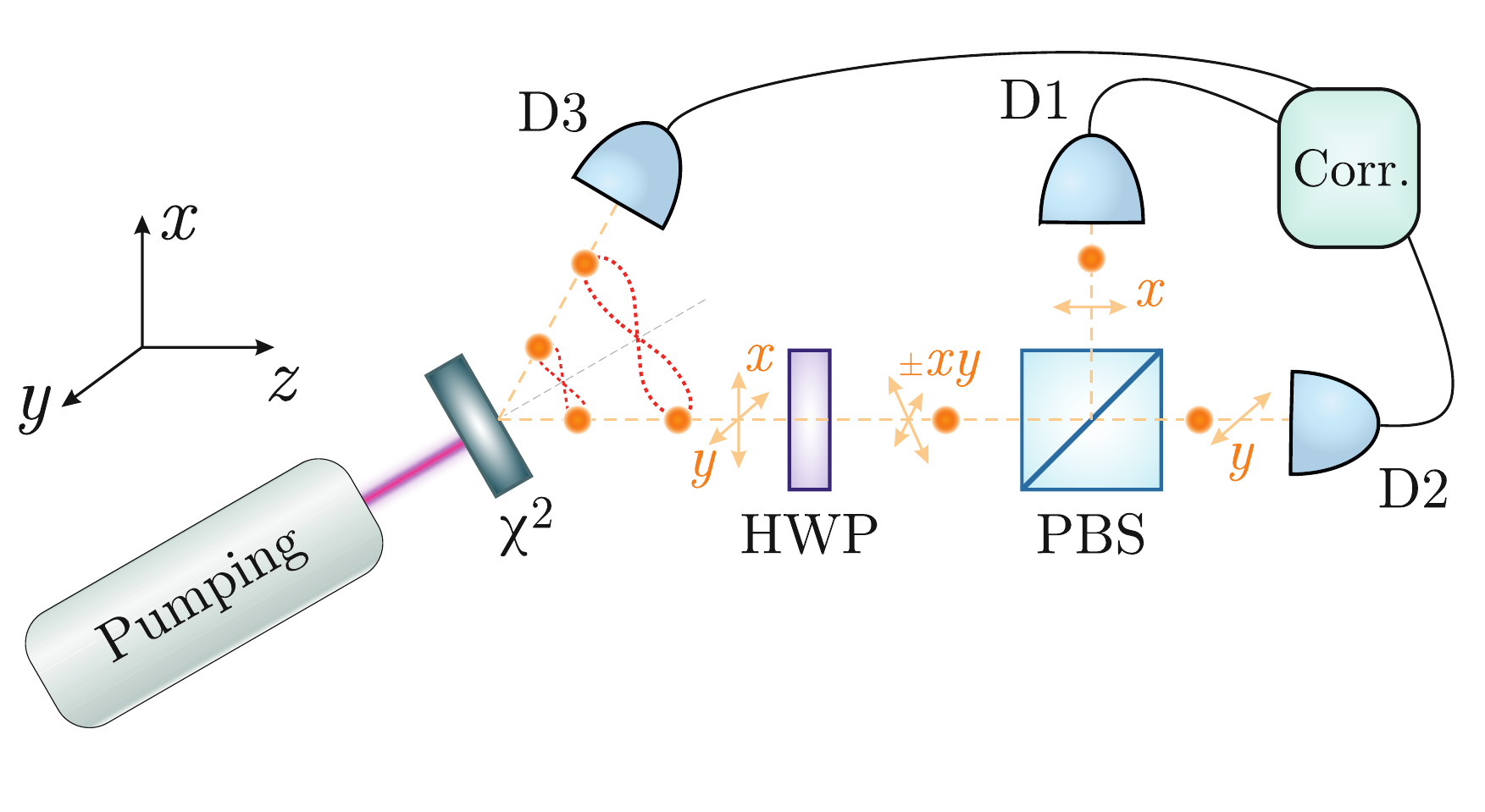}
    \caption{The heralded (genuine) single-photon source of RNSs: pumping a nonlinear $\chi^{(2)}$ crystal with laser radiation generates correlated photon-pairs in the entangled state $|\psi\rangle =2^{-1/2}(|H_1H_2\rangle +|V_1V_2\rangle)$. One photon from each pair is measured by the ''heralding'' detector (D3), signaling the existence of its partner. The polarization of this second, "heralded" photon is rotated to the diagonal basis using the half-wave plate (HWP). It is then directed to the polarizing beam splitter (PBS) where it has a fundamentally random probability of being measured by either detector (D1) or detector (D2), thus generating a random bit.}
    \label{fig2: quantum source}
    \end{figure}

In the second experimental setup (Fig.~\ref{fig2: quantum source}) similar to \cite{hai-qiang_random_2004,shafi_multi-bit_2023}, a RNS is generated using correlated photon-pairs.
The process is initiated by pumping a non-linear Type-2 BBO crystal (NLCQ5) with a stable $405$ nm laser diode (WSLD-405-020m-1). The pump laser's vertically polarized light undergoes spontaneous parametric down-conversion (SPDC) within the crystal, coherently generating pairs of photons at $810$ nm  in the entangled state $|\psi\rangle =2^{-1/2}(|H_1H_2\rangle +|V_1V_2\rangle)$, where $|H_i\rangle$ and $|V_i\rangle$ denotes horizontal and verical polarization of the $i$th photon. The detection of one photon from a pair at the detector D3 serves a dual purpose: it ''heralds'' the presence of its partner photon in the main channel, and simultaneously triggers a registration time-window to capture that single photon's arrival. The polarization state of this photon is transformed into a diagonal state using a half-wave plate (HWP). Thus, it has an equal probability of being reflected or transmitted by the polarizing beam splitter. A subsequent detection of the photon at the detector D1 is recorded as a binary ''0'', while a detection at the detector D2 is recorded as a ''1'', thereby producing the random sequence.
The primary advantage of this approach is its robustness to fluctuations in the pumping laser, which is achieved by configuring the optical setup to operate in the single-photon regime (further details are available in Appendix \ref{sec: app: hom effect}). While the system's performance remains sensitive to the ellipticity of the pump laser's polarization,  -- a factor that directly shapes the probability of generating specific quantum states, -- this does not require ongoing calibration. The setup needs only a single, initial alignment to synchronize the crystal's axes with the laser diode's polarization plane. This initial configuration, combined with the pump diode's low intrinsic ellipticity, allows for the generation of extended RNSs (over $5 \times 10^7$ bits) without a feedback loop. The primary trade-off is a lower generation rate, caused by the nonlinear crystal's SPDC efficiency.
The NIST test results (see Fig.~\ref{fig4: individual test results}b and Fig.~\ref{fig5: aggregated test results}b) confirm the efficiency of the heralded single-photon source scheme in generating high-quality random sequences. However, its p-value distributions show minor deficiencies. We attribute this primarily to the scheme's low generation rate. The extended data collection period made the system susceptible to laboratory environmental fluctuations, such as variations in temperature and humidity. These fluctuations altered the characteristics of the nonlinear crystal \cite{Takachiho-14}, thereby reducing the entropy of the randomness source.
In conclusion, we would like to note that the experimental setup yielded another significant result: the subsequences selected from the primary random string demonstrated remarkable efficiency. Specifically, when using the heralded single-photon source, the test subsequences exhibited a higher success rate than those from any other random number source tested. The significant randomness inherent in these strings made them an ideal benchmark RNS for a technique designed to enhance randomness by mixing sequences from two sources.
\subsection{A hybrid source}
\begin{figure}[htbp]
    \centering
    \includegraphics[width=\columnwidth]{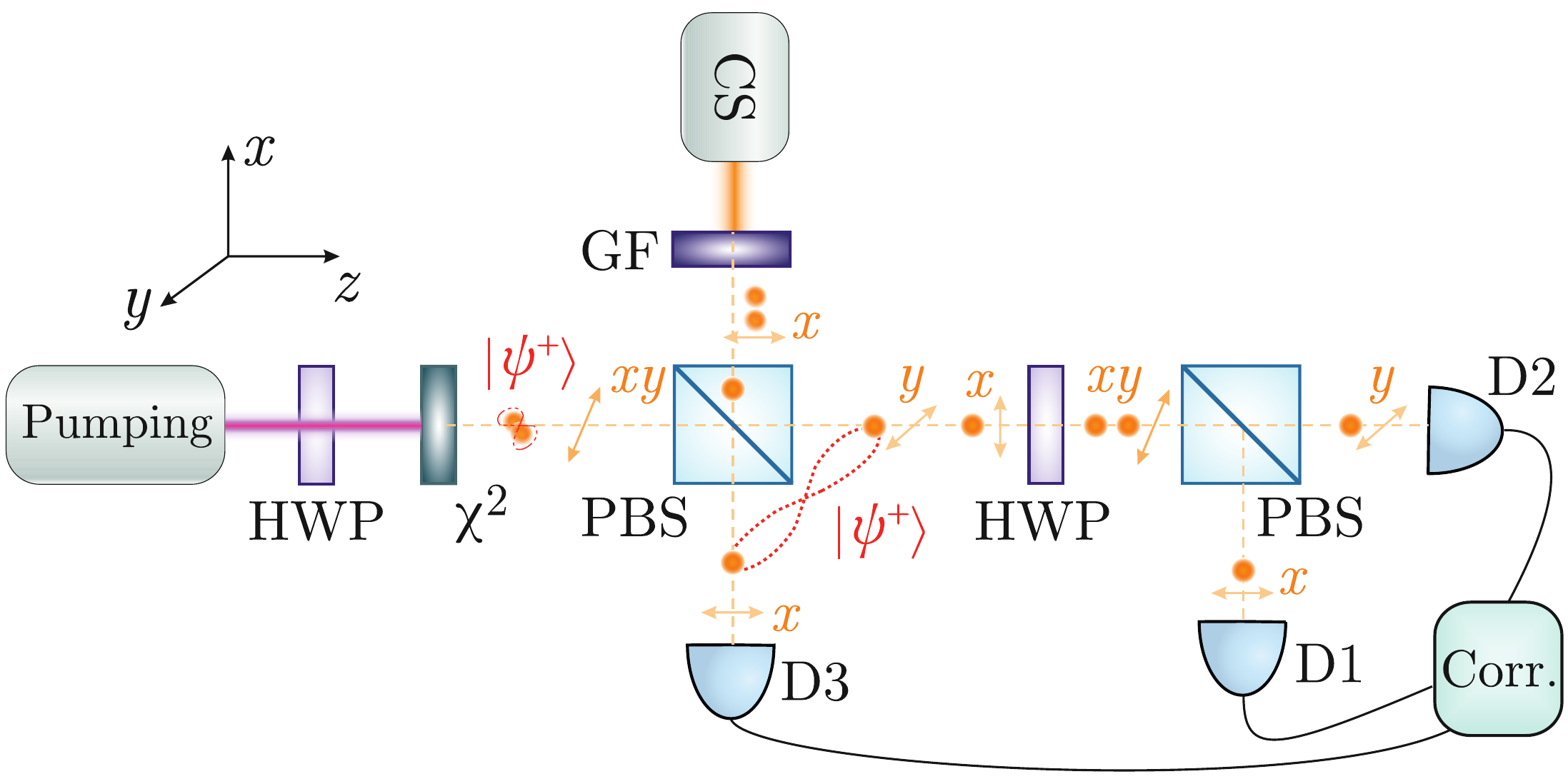}
    \caption{The hybrid source of RNSs: the $\chi^2$-crystal generates the Bell state $|\psi^{+}\rangle=2^{-1/2}(|H_1V_2\rangle+|V_1H_2\rangle)$, representing a pair of photons in a superposition of horizontal (H) and vertical (V) polarizations.  One photon from the pair is reflected and strikes the detector 3 (D3). Simultaneously, the entire output of the coherent source (CS) is reflected by the same PBS and mixed with the heralded photons. After polarization rotation with the HWP, the combined radiation is then directed to the second PBS and then measured by either detector 1 (D1) or detector 2 (D2), thus generating a random bit sequence.}
    \label{Fig3: hybrid source}
    \end{figure}

We propose and examine a novel hybrid RNG architecture (Fig.~\ref{Fig3: hybrid source}) that merges the semiclassical and quantum approaches considered above. This system uses the weakened coherent quasi-single-photon source as its primary emitter. To amplify quantum randomness, its output is combined with photons from the heralded single-photon source. The system is calibrated such that each heralded photon count is accompanied by approximately 20 counts from the coherent source. The heart of this scheme is the D3 detector, which plays a critical role in balancing the contributions from the single heralded photons and the coherent radiation. 
The process begins within the nonlinear crystal, which is configured for orthogonal coupling. When pumped by a laser with the diagonal polarization, the crystal generates the maximally entangled Bell states $|\psi^{+}\rangle=2^{-1/2}(|H_1V_2\rangle+|V_1H_2\rangle)$, representing a pair of photons in a superposition of horizontal (H) and vertical (V) polarizations.
This entangled pair is then directed to the first polarizing beam splitter (PBS). One photon from the pair is reflected and strikes the D3 detector. This "heralding" event signals that a single photon is now present in the main registration channel. Simultaneously, the entire output of the coherent source is reflected by the same PBS and  co-propagates with  the heralded single photons. To ensure an equal probability of detection, a half-wave plate (HWP) in the main channel is precisely rotated to equalize the photon count rates between the first D1 and the second D2 detectors. The recording of the RNS follows the same  procedure as previously described. We emphasize that the basis rotation with the second HWP is crucial here. After the first PBS, the photons from the coherent and heralded sources are orthogonally polarized (along the x- and y-axes in Fig.~\ref{Fig3: hybrid source}, respectively). To ensure that any photon has a 50/50 chance of being reflected or transmitted at the second PBS, one has to rotate them into the diagonal basis.
The proposed scheme is designed to eliminate the key limitations of the previous two. The primary advantage is the integration of the coherent source, which significantly increases the rate of RNSs generation.  This accelerated data collection mitigates the detrimental effects of low-frequency environmental drift on circuit components, thereby preserving the integrity of the random signal. The second key enhancement is the implementation of the heralded quantum architecture, which rise randomness by exploiting the fundamentally probabilistic nature of single-photon detection events. As demonstrated by the p-value distribution analysis (see Fig.~\ref{fig4: individual test results}c and Fig.~\ref{fig5: aggregated test results}c), this hybrid approach effectively amplifies the inherent randomness of the coherent source.
We would like to note, that the nature of the randomness is intrinsically linked to the hybrid architecture of its source. Essentially, by adding individual genuine random bits obtained from the heralded single-photon source to the pseudorandom number sequence from the quasi-signle-photon source, we eliminate residual correlations caused by the presence of multiphoton processes and the ellipticity of the semiclassical source's polarization. Moreover, the more such photons we add, the more the hybrid source behaves like the genuine RNG, which is confirmed by the subsequent analysis  carried out in the next section. 
\section{Results: Benchmarking of the  RNG(s) under consideration} \label{sec: benchmarking}
The raw data produced by physical RNGs often contain biases and correlations, which can be either inherent to ocassional coincidence or indicative of systematic errors. To reliably identify the last, extensive RNSs are required. For our study, we have generated 40 million bits from each source, including the software-based (''Python'') source for comparison. As one can see further, this sample size is sufficient for evaluating all systematic errors. In this section, we present a comprehensive analysis of the raw outputs from all RNGs and, for comparison, the results obtained with the post-processed algorithms. The post-processing techniques applied were randomness extraction via the von Neumann and V.F. Babkin methods (Section \ref{subsec: Benchmarking extractors})  and digital mixing with genuine random bits (Section \ref{subsec: Benchmarking mixing}).
\subsection{Evaluating RNGs performance with the NIST Tests: Raw Data vs. the Alchemy of Randomness Extraction}\label{subsec: Benchmarking extractors}
\begin{figure*}[!htb]
    \centering
    \includegraphics[width=\textwidth]{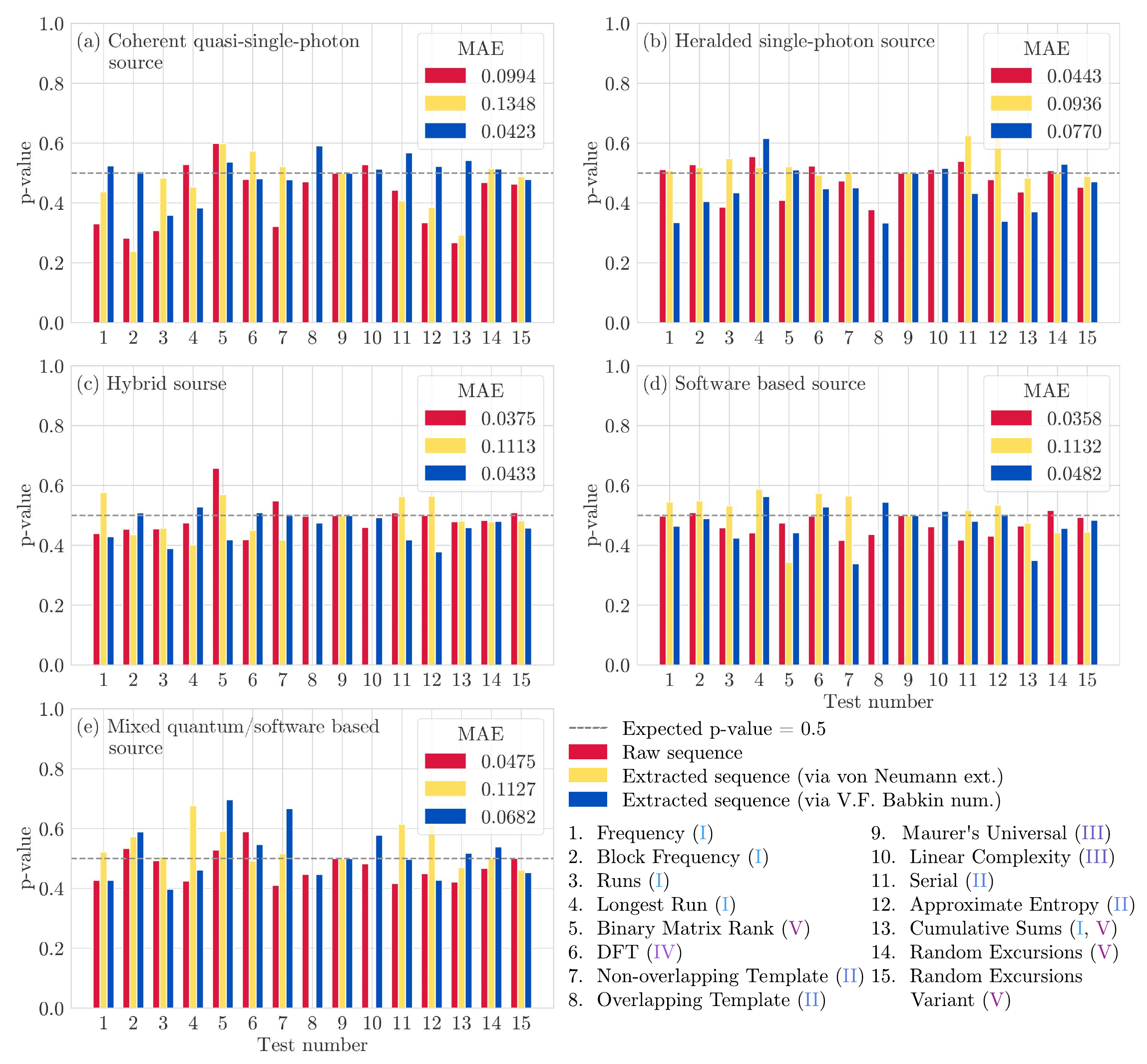}
    \caption{Median p-values of the NIST Statistical Test Suite for the five RNSs sources: a) Coherent single-photon source, b) Heralded single-photon source, c) Hybrid source, d) Software-based source, and e) Mixed quantum/software-based source. The tests are grouped into five categories for clarity: (I) ''Balance'', (II) ''Template'', (III) ''Complexity'', (IV) ''Spectral'', and (V) ''Structural''. The dashed line indicates the expected p-value =$0.5$.}
    \label{fig4: individual test results}
\end{figure*}
The randomness of the experimental dataset was evaluated using the NIST Statistical Test Suite \cite{rukhin_statistical_2010}, which comprises 15 distinct tests. To simplify their description, these tests can be grouped into five non-official categories based on the cryptographic properties they examine. The I'st category ''Balance'' (Tests 1, 2, 3, 4, 13): this category assesses the proportion of zeros and ones, as well as the occurrence of anomalous sequences of identical bits (runs).The II'nd category ''Template'' (Tests 7, 8, 11, 12): these tests search for repeating structures or a lack of diversity in substrings by looking for predefined patterns. The III'rd category ''Complexity'' (Tests 9, 10): this group evaluates linear complexity and the sequence's susceptibility to compression, which is indicative of redundancy \footnote{Here, the term ''compressibility'' refers to how compactly a sequence can be represented without loss of information. This concept is closely related to the sequence's informational entropy and data redundancy. Accordingly, an easily compressible sequence contains repeating patterns, whereas a truly random sequence is incompressible, as it lacks regularities and every bit conveys information. Maurer's Universal Statistical Test (9) attempts to compress the input sequence by exploiting repeating patterns in the initialization block.}. The IV'th category ''Spectral'' (Test 6): the Discrete Fourier Transform Test detects periodic patterns and spectral anomalies within the sequence. The V'th category ''Structural'' (Tests 5, 13, 14, 15): these tests evaluate the behavior of random walks (excursions) and the ranks of binary matrices. We note that The Cumulative Sums Test (13) occupies a unique position, bridging the ''Structural'' and ''Balance'' categories due to its hybrid methodology combining frequency and Markov chain analyses. We emphasize that these categories are not defined by NIST but were adopted for descriptive clarity.
For each source under consideration, we generated $20$ sequences of $2\times10^6$ bits each, resulting in a total of $40\times10^6$ bits per source. In addition to analyze the quality of raw sequences, we processed them using two entropy extractors: the von Neumann extractor and the stream numbering of Bernoulli sequences proposed by V.F. Babkin \cite{babkin_method_1971, arbekov_extraction_2021, arbekov_quantum_2024}. 

Figure \ref{fig4: individual test results} provides the obtained median p-values of the NIST statistical test suite for the original and post-processed data. The graphs include a black dashed line at expected p-value$\,=0.5$ to represent the ideal result for a random sequence. It is important to note that Maurer's Universal Statistical Test (9) has an expected p-value of $1$. To ensure consistent scaling across the graph, the output of test (9) was normalized to a value of 2. This normalization factor was incorporated into the calculation of the Mean Absolute Error (MAE), the numerical value of which is displayed on each plot. The analysis of the results is detailed below.
\subsubsection{The coherent quasi-single-photon source (Fig.~\ref{fig4: individual test results}a)}
The raw data showed the MAE of $0.0994$. The primary source of this deviation was a significant imbalance in the proportion of ''0''s and ''1''s, attributed to the sensitivity of the optical polarizers during generation, where minute angle misalignment favored one detector. Significant deviations were also observed in the template tests $(7, 12)$, which can also be attributed to the underlying imbalance.
The application of the von Neumann extractor effectively restored the bit balance, but it also precipitated a significant increase in the MAE to $0.1348$. This degradation in performance is a known trade-off of the technique. The extractor's process of discarding bits drastically shortens the sequence, undermining the statistical power of tests $(7, 12)$ on the resulting data, especially for sequences shorter than $10^6$ bits. This trend is observed for all sources under consideration.
The V.F. Babkin's stream numbering  yielded the best overall results, reducing the MAE to $0.0423$. It effectively managed the balance issues without the severe length reduction of the von Neumann technique. A minor degradation in tests (3) and (4) was noted, likely because the algorithm can create new blocks of repeated bits, which slightly impacts certain balance tests.
\subsubsection{The heralded single-photon source (Fig. \ref{fig4: individual test results}b)}
The raw data sequence exhibited a minor deviation from the expected p-value distribution, with the MAE of $0.0443$. The poorest performance was observed in tests $(3, 5, 8)$, signaling anomalies in the bit alternation structure.
Processing with the von Neumann extractor increased the MAE to $0.0936$. As with other sources, this aggregate increase is largely a consequence of failures in tests $(8, 10)$, which are sensitive to the shortened sequence length $(<10^6)$ post-extraction. Importantly, the majority of tests showed enhanced results. The successful passage of tests $(3, 5)$ with near-expected p-values provides evidence that initial correlations were effectively mitigated.
The application of the V.F. Babkin's stream numbering yielded an intermediate MAE of $0.0770$. This method produced divergent effects: it improved outcomes for structural group tests but caused a performance drop in balance and template group tests. This dichotomy implies that the algorithm may improve certain structural properties at the cost of degrading local bit distribution and alternation.
\subsubsection{The hybrid source (Fig.~\ref{fig4: individual test results}c)}
Among the considered physical RNGs, the hybrid source demonstrated the best performance for the raw sequence, with the MAE of just $0.0375$. The poorest results were in the spectral and structural group tests $(5, 6)$, likely due to the mixing process employed.
Application of the von Neumann extractor increased the MAE to $0.1348$. As stated above, this degradation is primarily attributed to the significantly reduced length of the extracted sequence $<10^6$, which caused interruptions in tests $(8, 10)$. While the extractor improved the results for tests $(5, 6)$, it either worsened the p-values for other tests or left them similar to the raw sequence.
In contrast, the V.F. Babkin's stream numbering yielded  the MAE of $0.0443$. It improved performance across a broad range of tests, including balance $(2, 4)$, structural $(5)$, spectral $(6)$, template $(7)$, and complexity $(10)$, highlighting its versatility. However, the algorithm's design led to a downturn in the results for template tests $(8, 11, 12)$.
\subsubsection{The software based source (Fig.~\ref{fig4: individual test results}d)}
For benchmarking the physical RNGs, we included pseudorandom sequences from Python's ''Secrets'' library in our analysis. The ''raw'' sequence achieved the lowest MAE of $0.0358$, outperforming all physical sources under consideration. This superior performance is expected, as cryptographically secure pseudorandom number generators are explicitly designed to pass stringent statistical tests like the NIST suite.
However, this performance is fragile. Applying the von Neumann extractor increased the MAE to 0.1132, primarily by disrupting several structural tests $(5, 8, 10)$. This suggests that the extractor disrupted the carefully engineered structure of the original sequence. Paradoxically, the extractor improved performance on some template-based tests. This implies that the original "raw" sequence contained subtle, algorithmically-generated patterns that were successfully removed.
The V.F. Babkin's stream numbering also increased the MAE, though less drastically, to $0.0482$. The decline was primarily observed in tests measuring bit-balance and in test $(7)$. We hypothesize that the method can alter the ratio of ''0''s and ''1''s, thus failing balance tests, while simultaneously creating new, repetitive patterns that are detected by test $(7)$.
\subsubsection{The mixed quantum/software based source (Fig.~ \ref{fig4: individual test results}e)}
Driven by curiosity, we also evaluated a mixed physical-software based RNG. This source was ''constructed'' by taking a pseudorandom sequence from the Python's ''Secrets'' library (d) and replacing every 20th bit with a bit from the heralded single-photon source (b).
The "raw" sequenceof the mixed source  had the MAE of $0.0475$. This decrease of performance over the pure software based source (d) is likely attributable to the disruption of the ''perfected'' algorithmic structure of the pseudorandom sequence.
Application of the von Neumann extractor increased the MAE to 0.1127. This degradation was primarily due to the failure of tests $(8, 10)$, but a significant decline was also observed in the balance test $(4)$ and the template tests $(11, 12)$, indicating a substantial loss of entropy in the post-processed sequence.
 Processing with the V.F. Babkin’s stream numbering method yielded the MAE of $0.0682$, which still exceeds the deviation of the raw sequence. Notable failures occurred in tests $(5, 7, 10)$, likely due to artifacts introduced by the extraction method.
\subsubsection{The overall assessment of the RNGs performance}
\begin{figure*}[!htb]
    \centering
    \includegraphics[width=1\linewidth]{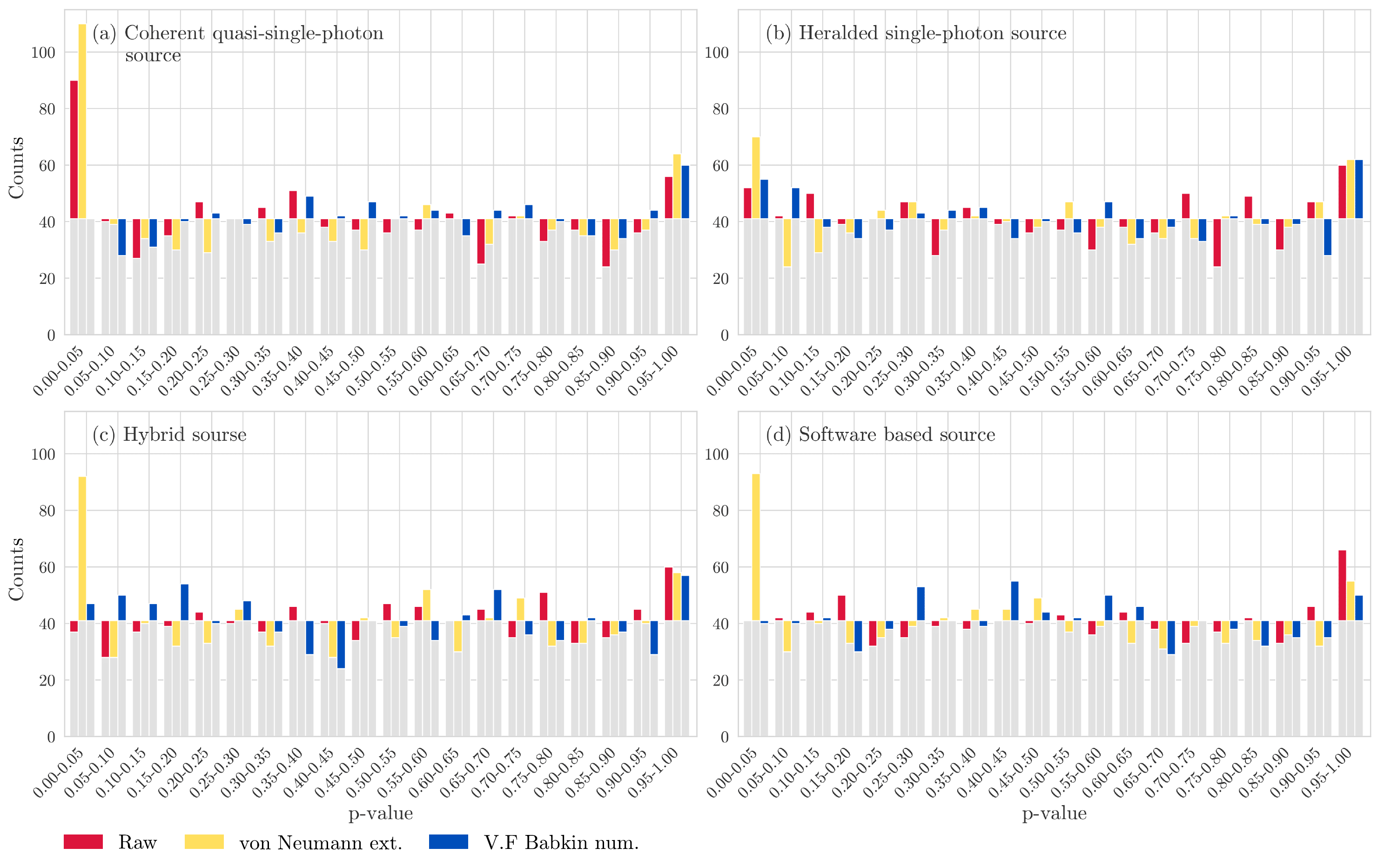}
    \caption{Comparison of p-value distributions obtained for raw and post-processed sequences for: a) Coherent single-photon source, b) Heralded single-photon source, c) Hybrid source, and d) Software-based source. The p-value intervals are plotted on the x-axis; the count of p-values falling within the specified intervals is indicated on the y-axis. In case of the uniform distribution, the Count value for all intervals is $42$.}
    \label{fig5: aggregated test results}
\end{figure*}

While the analysis above identified asymmetry in the p-value distributions of the studied sources, it is important to note its key limitation: the median p-value alone is an insufficient metric for confirming randomness. p-value=$0.5$ is indeed characterize a uniform distribution, which suggests random signals. However, the same median value can also be produced by other symmetric distributions -- for instance, a normal distribution centered at $0.5$. Therefore, to assess the overall quality of the generated RNSs, we summed the p-value distributions from all NIST statistical tests and calculated the average distribution across them. This analysis was performed on a set of $20$ RNSs, each comprising $2\times10^6$ bits. Figure \ref{fig5: aggregated test results} presents the resulting distributions for all three physical RNGs under consideration as well as the software based source. Under the assumption of randomness, the distribution of p-values for the studied sequences is expected to be uniform \cite{klammer_statistical_2009}. The gray bars represent this theoretical uniform distribution for the given sample size, while the colored bars visualize the empirical deviation from it. For context, a p-value of $0.01$ is established as the default passing criterion.

\begin{figure}[h!]
    \centering
    \includegraphics[width=\columnwidth]{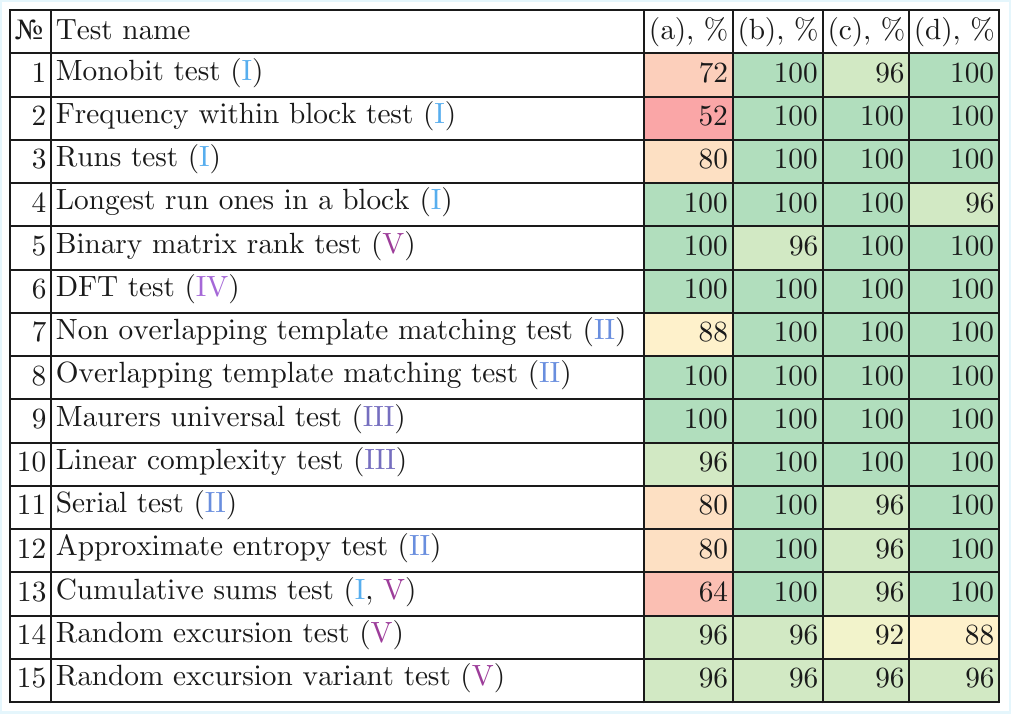}
    \caption{The percentage of tests passed by the subsequences for:  a) Coherent quasi-single-photon source, b) Heralded single-photon source, c) Hybrid source, and d) Software-based source. The tests are grouped into five categories for clarity: (I) ''Balance'', (II) ''Template'', (III) ''Complexity'', (IV) ''Spectral'', and (V) ''Structural''. The threshold for a passing p-value was set at the default of $0.01$.}
    \label{Fig: (table) a percentage of tests passe table}
    \end{figure}

For the coherent quasi-single-photon source (Fig.~\ref{fig5: aggregated test results}a), the raw data exhibits a significant deviation in the $[0.00-0.05)$ interval, indicating a high failure rate in statistical tests and unsatisfactory randomness. While applying the von Neumann extractor also resulted in numerous test failures, this was not due to poor randomness but rather the algorithm's high data demands, which truncated the sequence length below the threshold required for certain tests. In contrast, the application of the V.F. Babkin's stream numbering method corrected the initial bias and produced a significantly improved distribution.
The raw data from the heralded single-photon source (Fig.~\ref{fig5: aggregated test results}b) exhibits smaller deviations from a uniform distribution than source (a), indicating superior inherent randomness. However, applying the von Neumann post-processing method increases the number of failed statistical tests, mirroring the results from the first source. Despite this, the overall magnitude of the deviations remains consistent. The V.F. Babkin's stream numbering demonstrates a slight improvement over the original raw sequence.
Among the three generation schemes, the raw sequence from the hybrid source (Fig.~\ref{fig5: aggregated test results}c) yielded a distribution closest to the ideal uniform case. This enhancement stems from the mixing of photons from sources (a) and (b) -- a technique aligned with established methods for randomness extraction from multiple sources with different entropy levels \cite{nisan_extracting_1996}. This effect is explained by the additivity of entropy, which implies that a hybrid source possesses a level of randomness no lower than that of its components. Consequently, by combining sources with varying entropy levels, we effectively increase the overall entropy, producing a superior stream of random bits. The application of both the von Neumann extractor and the V.F. Babkin's stream numbering method degrades the p-value distribution. This indicates that for data sources of already high entropy, such post-processing is not only redundant but may even be counterproductive to statistical quality.
A similar pattern emerges for the  software based source (Fig.~\ref{fig5: aggregated test results}d): while the initial raw sequence demonstrates satisfactory randomness, this quality degrades following post-processing. These findings underscore the critical need for adaptive randomness extraction methods, tailored to a source's specific characteristics.
 The fundamental principle of randomness dictates that any subsequence from a truly random string must itself be random and, therefore, pass statistical tests -- provided its length meets the minimum requirement \cite{arbekov_extraction_2021}. Consequently, the overall randomness of the whole sample can be gauged by the proportion of its subsequences that pass these tests. As summarized in Fig. \ref{Fig: (table) a percentage of tests passe table}, which details the results for subsequences from three physical RNGs under consideration and the software based source, a higher pass rate indicates greater randomness. 
The results obtained reveals that subsequences from the heralded single-photon source consistently pass a high proportion of tests, confirming its quality. In contrast, the hybrid source performs slightly worse. Noteworthy, that the software generator despite its overall stronger performance demonstrated a significant weakness, achieving the poorest results of all sources in the Random Excursion test.

\subsection{Boosting Randomness:Digital Mixing  of Pseudorandom Sequences with Genuine Random Bits} \label{subsec: Benchmarking mixing}

Beyond the physical ''mixing bits'' by mixing photons from the semiclassical and quantum sources in the hybrid source, digital bit mixing presents a compelling alternative. We estimated the potential to enhance the randomness of the coherent quasi-single-photon source (a) by digitally replacing every $i$th bit in its output with a corresponding bit from the heralded quantum source (b). The randomness of the resulting sequences was assessed using the NIST statistical test suite. For benchmark purposes, we performed an identical digital mixing procedure between the software-based source (d) and the heralded source (b) and, analyzing the resulting RNSs of equal length.

\begin{figure*}[!htb]
    \centering
    \includegraphics[width=0.95\linewidth]{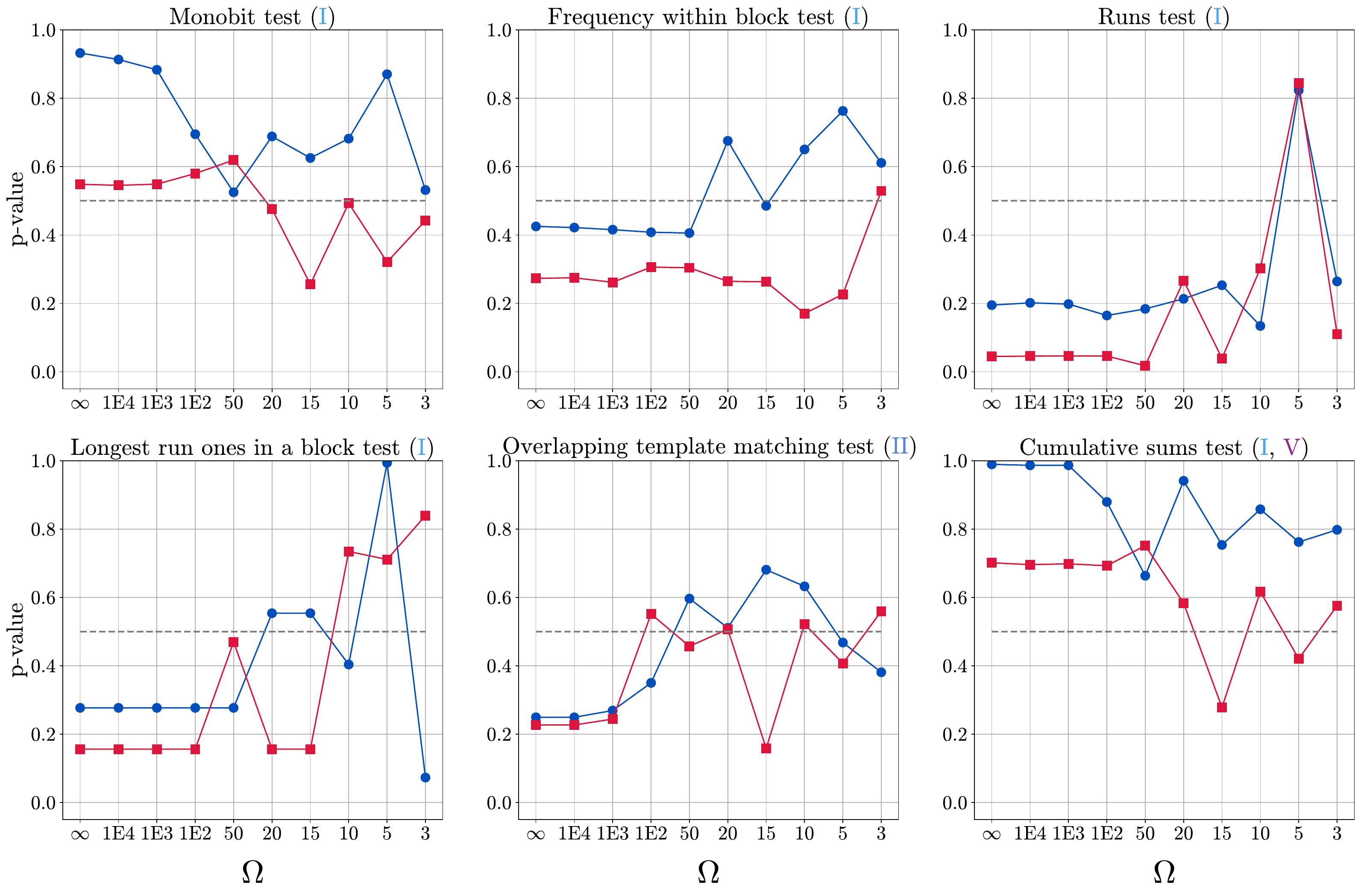}
    \caption{Randomness quality as a function of mixing frequency for: $\square$ (red) -- balanced coherent quasi-single photon source,  $\circ$ (blue) -- software-based ''Python'' source. The dashed line indicates the expected p-value =$0.5$. The tests represent following groups: (I) ''Balance'', (II) ''Template'', and (V) ''Structural''. $\Omega$ denotes the proportion (or ''period'') of pseudorandom bits to genuine random bits in the mixed random number stream} .
    \label{fig7:mixing results}
\end{figure*}

The most noticeable enhancement was observed across six statistical tests of the NIST suite. The part of these tests measures bit balance, including the monobit, frequency within a block, runs, longest run ones in a block,  and  cumulative sums tests. The rest assesses the prevalence and variety of patterns. As illustrated in Figure \ref{fig7:mixing results}, the mixing strategy significantly improved the sequence's statistical properties. The results of all test are provided in Appendix \ref{sec: app: mixing sequences}. 
\begin{figure*}[!htb]
    \centering
    \includegraphics[width=\textwidth]{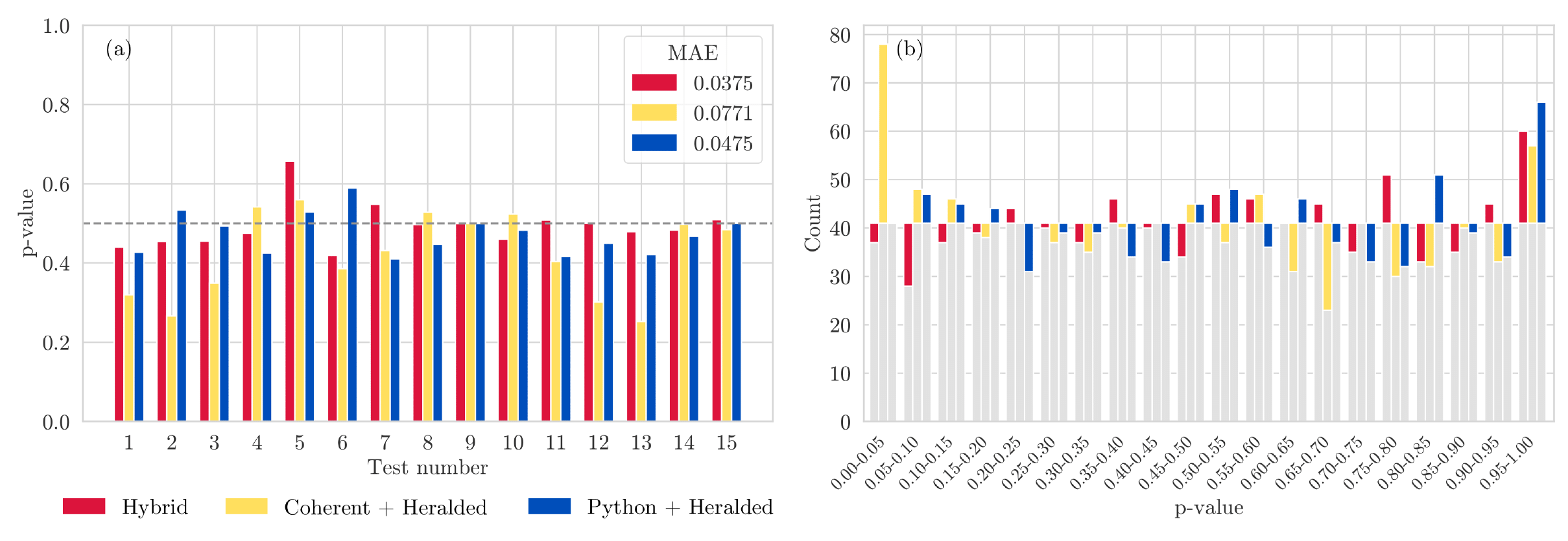}
    \caption{a) Median p-values from the NIST Statistical Test Suite for randomness quality assessment. Results are shown for the hybrid source and for digitally mixed bits from two combinations: the coherent quasi-single-photon source with the heralded source, and the software-based source with the heralded source.
    b) The averaged distribution of these p-values across the entire NIST test suite for all characterized sources.}
    \label{fig8: comparison of physical and digital mixing}
\end{figure*}
Our analysis reveals that for each test from the NIST suite there is a  unique optimal frequency for mixing with true random numbers. Therefore, achieving a balance between mixing frequency and the resulting p-value for all tests is critical.  Accordingly, we chose a ratio of $1:20$ between truly random and pseudorandom bits for the comparison of the digital mixing method with the hybrid source (c).

Fig.~\ref{fig8: comparison of physical and digital mixing} presents the individual test results, along with the averaged p-number distribution for RNSs generated by the physical hybrid source and the digital bit-mixing method. We emphasize that  other post-processing techniques were not applied. As one can see, the physical source maintained a slight advantage over both digital approaches, including the scenario where the digital mixing used the software-based source as its primary input. Therefore, raw RNSs generated by the proposed hybrid source are suitable for direct application in advanced optical experiments, eliminating the need for any supplementary digital post-processing.

\section{Conclusion} \label{sec: conclusion}
In our work, we have presented the systematic, comparative analysis of three physical RNG sources -- semiclassical, quantum, and the novel hybrid source -- implemented on a single reconfigurable hardware optical platform. This unified approach enabled a definitive assessment of their performance trade-offs: while the semiclassical RNG maximizes speed, and the quantum RNG delivers superior quality, the hybrid source effectively combines relatively high speed with robust randomness, demonstrating an optimal balance.
For benchmarking, we compared all three physical-based RNGs against the cryptographically secure pseudorandom source from the Python's ''Secrets'' library employing the NIST statistical test suite. Our analysis reveals that the raw sequences from the proposed hybrid RNG, not only match the statistical quality of the heralded and pseudorandom software-based source but even exceed it in several key metrics. Notably, the raw output from the hybrid RNG exhibited greater randomness than sequences refined by established post-processing techniques, including the von Neumann extractor and V.F. Babkin's stream numbering algorithm.
A key comparison was made between the inherent physical mixing of bits in the proposed hybrid source and software-based digital mixing. The results of the tests confirmed that the hybrid source maintains a certain advantage in performance, highlighting the benefit of leveraging physical processes for entropy generation.

Thus, following initial calibration, this hybrid source generates sequences with quality approaching that of the heralded and the software-based sources, and its straightforward design ensures direct compatibility with various quantum optical schemes in experiment.

Additionally, we note that the RNS generation architectures presented in the work are highly versatile and easily adapted to diverse experimental setups, making them suitable for any optical application requiring pseudorandom or genuine random sequences.  Their design is compatible with miniaturization and fabrication according to integrated optics standards \cite{integral_spdc}, thereby extending their potential application to fields such as fiber-optic communication and quantum key distribution \cite{comm_rng}.
\appendix
\section{
The stream numbering of Bernoulli sequences proposed by V.F. Babkin} \label{sec: app: V.F. Babkin's extractor}
\begin{figure*}[!htb]
    \centering
    \includegraphics[width=\textwidth]{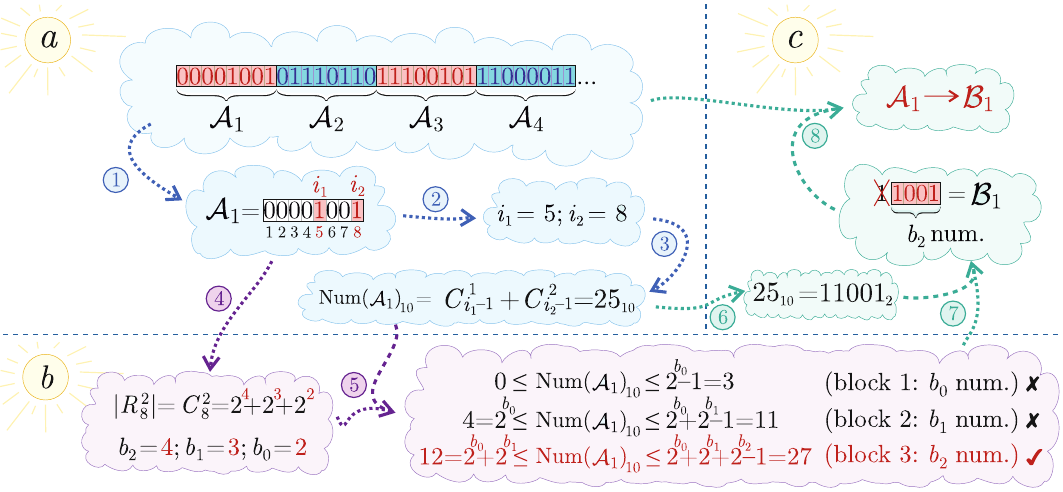}
    \caption{The example of the stream numbering proposed by V.F. Babkin. a) Step 1: a Bernoulli sequence $\mathcal{A}_1\mathcal{A}_2\mathcal{A}_3\mathcal{A}_4...$ is divided into subsequences of length $n=8$. Step 2: in each subsequence (e.g. $\mathcal{A}_1$), the position of the ''1''s is identified. Step 3: $\mathcal{A}_1$ is assigned a number $\text{Num}(\mathcal{A}_1)$ in accordance with  \eqref{eq: ap: 1 number}. b) Step 4: the cardinality $|R_n^k|$ of the set of all possible mutual permutations of $k=2$ ''1''s in $n=8$ positions is calculated. Step 5: employing the binary form \eqref{eq: ap: cardinality} of $|R_n^k|$ and inequality \eqref{eq: ap: block_inequality}, the belonging of $\text{Num}(\mathcal{A}_1)$ to specific block is identified. c) Step 6: The decimal form $\text{Num}(\mathcal{A}_1)$ is converted to binary  \eqref{eq: ap: num_binary_form}. Step 7: since $\text{Num}(\mathcal{A}_1)_{10}$ belongs to block 3, the new subsequence $\mathcal{B}_1$ contains the last $b_2$ numbers of the binary form $\text{Num}(\mathcal{A}_1)_2$. Step 8: $\mathcal{A}_1$ is replaced by $\mathcal{B}_1$.}
    \label{Fig: numbering}
    \end{figure*}

For the sake of completeness, the stream numbering of Bernoulli sequences proposed by V.F. Babkin \cite{babkin_method_1971, arbekov_extraction_2021, arbekov_quantum_2024} is reproduced herein. The method is comprised the following  steps:
\begin{enumerate}
    \item A generated Bernoulli subsequence $\mathcal{A}_1$ of length $n$ with $k$ ''1''s and $(n-k)$ ''0''s is assigned a number in accordance with the following expression: 
    \begin{align}
        \text{Num}(\mathcal{A}_1)_{10} =  \sum_{m=1}^k C_{i_m-1}^m, \label{eq: ap: 1 number}
    \end{align}
    where $\{i_1,i_2,...,i_k\}$ are the positions of ''1''s in $\mathcal{A}_1$, and $\{C_{i_m-1}^{m}\}$ are binomial coefficients. 
    \item The received number $\text{Num}(\mathcal{A}_1)_{10}$ is placed in one of the special blocks. In order to determine the total number of such blocks and the belonging of $\text{Num}(\mathcal{A}_1)$ to a specific block, it is necessary to find the total number of all possible unique permutations of $k$ ''1'' and $(n-k)$ ''0'', i.e. the cardinality of the set $R_n^k$: $|R_n^k|=\text{card}(R_n^k) =C_n^k$. Next, cardinality $|R_n^k|$ must be represented in a binary form or, alternatively, as a sum of powers of two:
    \begin{align}
        |R_n^k| =\sum_{i=1}^{m}2^{b_i}\label{eq: ap: cardinality},
    \end{align}
    where $b_m>b_{m-1}>...>b_1$. Thus, there are $m$ different blocks. The $1$st block includes numbers from 0 to $(2^{b_1}-1)$. The 2nd block comprises numbers from $2^{b_1}$ to $(2^{b_1}+2^{b_2}-1)$, and so on. To ascertain the belonging of $\text{Num}(\mathcal{A}_1)$ to the $k$th block, it is necessary to determine whether it satisfies the corresponding inequality:
    \begin{align}
        &\sum_{i=1}^{k-1}2^{b_i}  \leq \text{Num}(\mathcal{A}_1)_{10}\leq \sum_{i=1}^{k}2^{b_i}-1. \label{eq: ap: block_inequality}
    \end{align}
    \item The resulting number, expressed in decimal notation, is then converted to binary form:
    \begin{align}
        &\text{Num}(\mathcal{A}_1)_{10} \rightarrow \text{Num}(\mathcal{A}_1)_{2} =\varepsilon_{r_m}\varepsilon_{r_{m-1}}...\varepsilon_{1}\varepsilon_{0} ,\label{eq: ap: num_binary_form}
    \end{align}
where $\varepsilon_{r_m}\varepsilon_{r_{m-1}}...\varepsilon_{1}\varepsilon_{0} = \sum_{i=0}^m \varepsilon_{r_i} 2^{r_i}$, and $\forall i\;  \varepsilon_{i}\in\{0,1\}$.     Subsequently, from the binary form $\text{Num}(\mathcal{A}_1)_2$, a new subsequence, designated $\mathcal{B}_1$, is formed, which includes the last $b_k$ numbers of $\text{Num}(\mathcal{A}_1)_2$, where $k$ is the number of the block to which $\text{Num}(\mathcal{A}_1)_{10}$ belongs, i.e. $\mathcal{B}_1= \varepsilon_{b_k}\varepsilon_{b_{k-1}}...\varepsilon_{1}\varepsilon_{0}$. It should be noted that $\mathcal{B}_1$ can only partially coincide with $\text{Num}(\mathcal{A}_1)_{2}$, i.e. $\mathcal{B}_1$ can include fewer ''0''s and ''1''s than $\text{Num}(\mathcal{A}_1)_{2}$.
  \item This process is then repeated for subsequences $\mathcal{A}_2$, $\mathcal{A}_3$, and etc. Ultimately, the sequence $\mathcal{A}_1\mathcal{A}_2\mathcal{A}_3...\mathcal{A}_N$ is replaced by the sequence $\mathcal{B}_1\mathcal{B}_2\mathcal{B}_3...\mathcal{B}_N$.    
\end{enumerate}

An example of the stream numbering method described above is illustrated in Fig.~\ref{Fig: numbering}.

\section{Operational performance of randomness extractors}
\label{sec: app: extractor performance}
The application of randomness extractors can enhance the statistical properties of raw data streams, leading to improved test results. However, the detailed analysis revealed that this process is not universally beneficial. For some sources, extraction is unnecessary if the raw sequence already demonstrates satisfactory randomness.
The operational principles of the chosen extractor are critical. For instance, the von Neumann extractor consistently improves balance tests, yet the MAE across all tests remains high. This is the direct consequence of the method's significant data loss: it discards at least $50$\% of the original bits, with considered sources (a)-(e) experiencing an average length reduction of $66$\%. This severe shortening sometimes caused tests $(8, 10)$ to fail entirely, as the resulting sequences were too short for the algorithms to execute. 
In contrast, the V.F. Babkin’s stream numbering introduces a different trade-off. While it causes significantly smaller bit losses than the von Neumann extractor and demonstrates considerable universality (as evidenced by its performance on the hybrid source (c)), it carries the risk of generating new, subtle patterns. Therefore, its application demands a cautious and informed approach. A key implementation challenge for the method is selecting the sample size $n$ -- the length of the sequence fragment processed at once.  The optimal choice is determined by balancing the computation time of binomial coefficients, the unavoidable bit losses, and the randomness quality of the extracted sequences, as assessed by statistical tests.
The computation of binomial coefficients can be approached through either exact or approximate methods. An exact calculation, which relies on the direct definition using factorials, possesses a time complexity of $O(n)$. However, in practical implementations, the computational cost often grows super-linearly due to the overhead of managing large integers in various programming languages. In contrast, an approximate computation leverages the Gamma function to provide a result in constant time of $O(1)$. While this approximation introduces a negligible margin of error and is inefficient for small values of $n$, it offers a substantial performance advantage for large-scale computations where $n \gtrsim 10^3$.  At the same time, for a fixed value of $n$, the optimal computational strategy is to precompute all relevant binomial coefficients and store them in a lookup table. This approach amortizes the initial calculation cost over numerous subsequent accesses, yielding a substantial reduction in execution time for algorithms that require repeated evaluations.
We note that, the V.F. Babkin’s stream numbering exhibits a trade-off: while larger values of $n$ reduce bit loss, they also increase computational complexity.  Although the approximate method alleviates time constraints, values of $n > 1000$ introduce significant limitations due to RAM and the programming language's capacity for handling large numbers. To assess the randomness quality of the extracted sequences, we tested values of $n \in \{100, 200, 300,  400\}$.  Our analysis demonstrated that $n = 300$ produced the smallest MAE. Consequently, this value was selected for the post-processing of the experimental data.

 \section{Photocount distribution of  the coherent quasi-single-photon  source} \label{sec: app: coherent source destribution}
\begin{figure}[htbp]
    \centering
    \includegraphics[width=1\columnwidth]{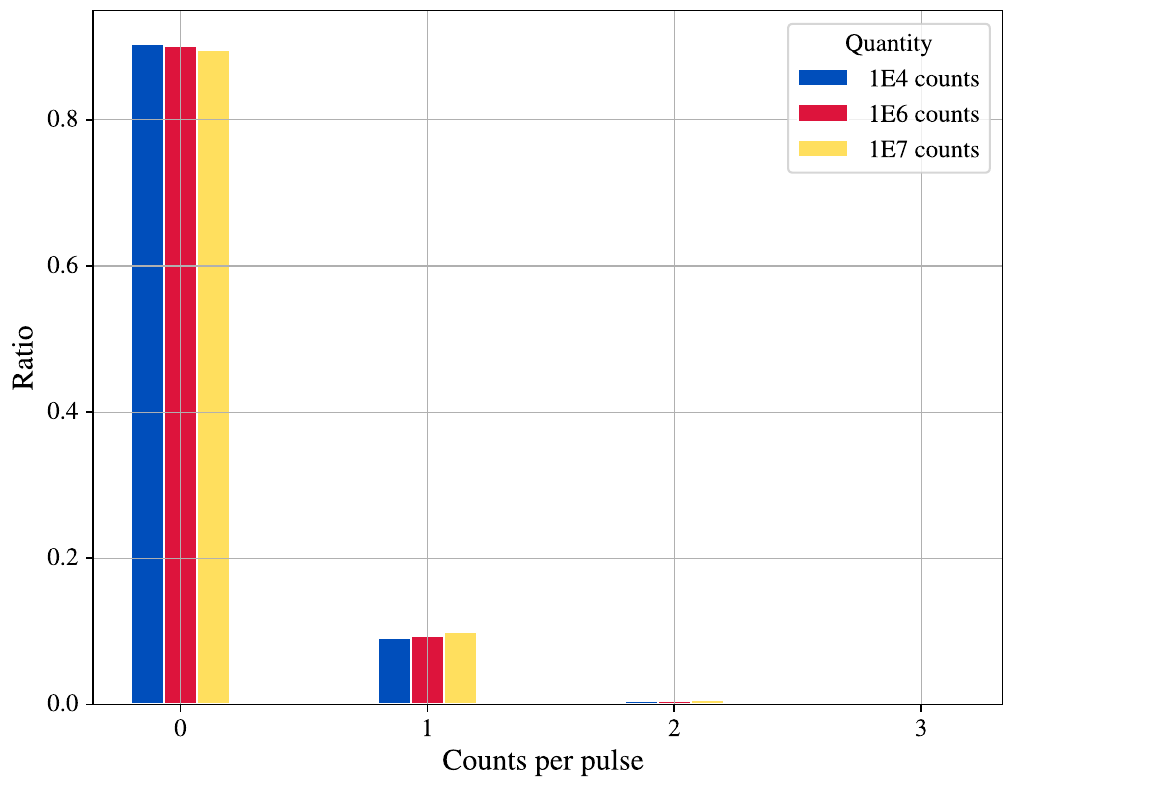}
    \caption{The distribution of photocount events for sequences of different lengths. The height of each histogram column represents the proportion of pulses that have a specific number of single-photon detector responses per correlator cycle, out of all the pulses.}
    \label{Fig: counts distribution}
    \end{figure}

A coherent light source exhibits the Poisson photocount distribution per pulse. In order to operate in the quasi-single photon regime, it is essential that the average photocount per pulse be approximately 0.1 or fewer. In the experiment, we employed  a continuous-wave laser and divided  its output into distinct pulses using an external pulse generator synchronized to a correlator. The frequency of the pulse generator was chosen such that the mean photocount per correlator cycle matched the desired value of 0.1. Fig.~\ref{Fig: counts distribution} displays a histogram of the photocount distribution over 3 independent sequences of different lengths ($10^4$, $10^6$ and $10^7$ counts). During the series of experiments, the pulse frequency of the generator was maintained constant, allowing the variation in photocounts to be attributed solely to fluctuations in laser intensity.

The average photocounts per pulse for the sequences obtained were $\lambda_{1e4}=0.095$, $\lambda_{1e6} = 0.105$, and $\lambda_{1e7} = 0.11$, respectively. While this deviation from the expected mean number is not significant for experiments involving random number generation, it indicates good stability of the laser source. Thus, fluctuations in polarization rather than intensity appear to have a more substantial impact on the randomness of generated sequences.

\section{Spectral width of the entangled bi-photon light field} \label{sec: app: hom effect}
In the experiment, we employed an orthogonal combination of two nonlinear Type 2 BBO crystals as the source of correlated photon pairs. The change in the generated polarization state was achieved by adjusting the pump beam waist. For the generation of the state with the same polarization $\big( \text{i.e.\;} |\psi\rangle =2^{-1/2}(|H_1H_2\rangle +|V_1V_2\rangle)\big)$, the waist was located on one of the two crystals; for the generation of  the state with different polarizations $\big(\text{i.e.\;} |\psi^{+}\rangle=2^{-1/2}(|H_1V_2\rangle+|V_1H_2\rangle)\;\big)$, it was located precisely between them.

One of the significant features of the biphoton field is its spectral bandwidth. Since single-photon detectors differs in their spectral sensitivity, this affects the efficiency of detecting single-photon emission. To measure the spectral bandwidth of the biphoton field, we employed the Hong-Ou-Mandel interference effect. We note that the very presence of interference will confirm the true quantum nature of the photon source.

\begin{figure}[htbp]
    \centering
    \includegraphics[width=1\columnwidth]{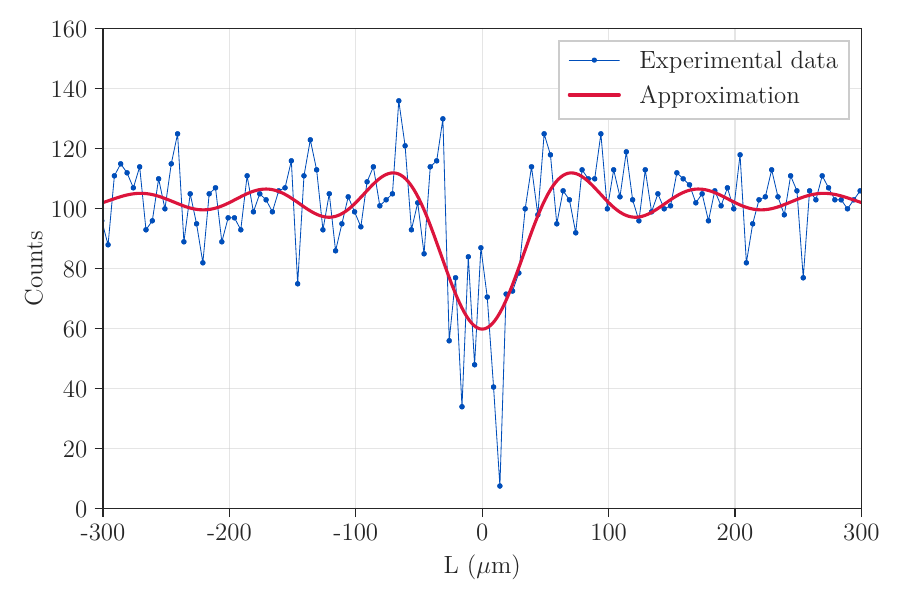}
    \caption{The dependence of the coincident count rates of single-photon detectors at the output of the two-photon interference experiment on the relative phase shift in the interferometer arms (in terms of length). The red line is the approximation of experimental data using function $f(L)=C(1+A\:\text{sinc}(L/w))$.}
    \label{Fig: HOM}
    \end{figure}

Pairs of entangled photons were sent through a Mach-Zehnder interferometer. A linear translator (a delay line) was used to change the relative phase shift on the output beamsplitter. After the beamsplitter, the radiation was detected by single-photon counters in the coincident counting mode. The dependence between the number of coincidences and the position of the linear translation stage is presented in Fig.~\ref{Fig: HOM}.

The graph in Fig.~\ref{Fig: HOM} has been centered such that the zero point of the linear scale corresponds to the absence of the relative phase shift. The presence of a significant discrepancy in the coincidence counts near the point $L=0$ demonstrates photon bunching in a nonlinear crystal. Probability of detection both photons in the same output channel is proportional to the function of the phase shift:

\begin{equation}
    P(L) \propto 1+\text{sinc}\Big(\delta\omega L/c\Big),
\end{equation}
where $\delta\omega$ -- spectral bandwidth in the frequency range, $L$ -- phase shift (in terms of length units). 

Approximating the experimental data with function of the form $f(L)=C(1+A\,\text{sinc}(\frac{L}{w}))$ ($C, A, w$ --- approximation constants) we found that $w=1.6\cdot10^{-5}$ m. Consequently, we obtained:
\begin{equation}
    \delta\omega=\frac{c}{w}=1.875\cdot 10^{13}~ \text{Hz},
\end{equation}
or in the wavelength range:
\begin{equation}
    \delta\lambda=\frac{1}{2\pi}\frac{\lambda_0^2}{w}  = 6.5~\text{nm},
\end{equation}
where $\lambda_0$ is the central wavelength of the biphoton field (810 nm). Therefore, the nonlinear crystal employed in the experiment exhibits a relatively narrow spectral bandwidth, which can be further narrowed by the use of narrow-bandpass spectral filters. 

The observed phenomenon of photon bunching and the measured bandwidth of the biphoton spectrum indicate that our source produces correlated the entangled states with high reliability.

\section{Mixing Pseudorandom Sequences with Genuine Random Bits} \label{sec: app: mixing sequences}

%[!htp]

%
%

\begin{figure*}
    \centering
    \includegraphics[width=0.95\textwidth]{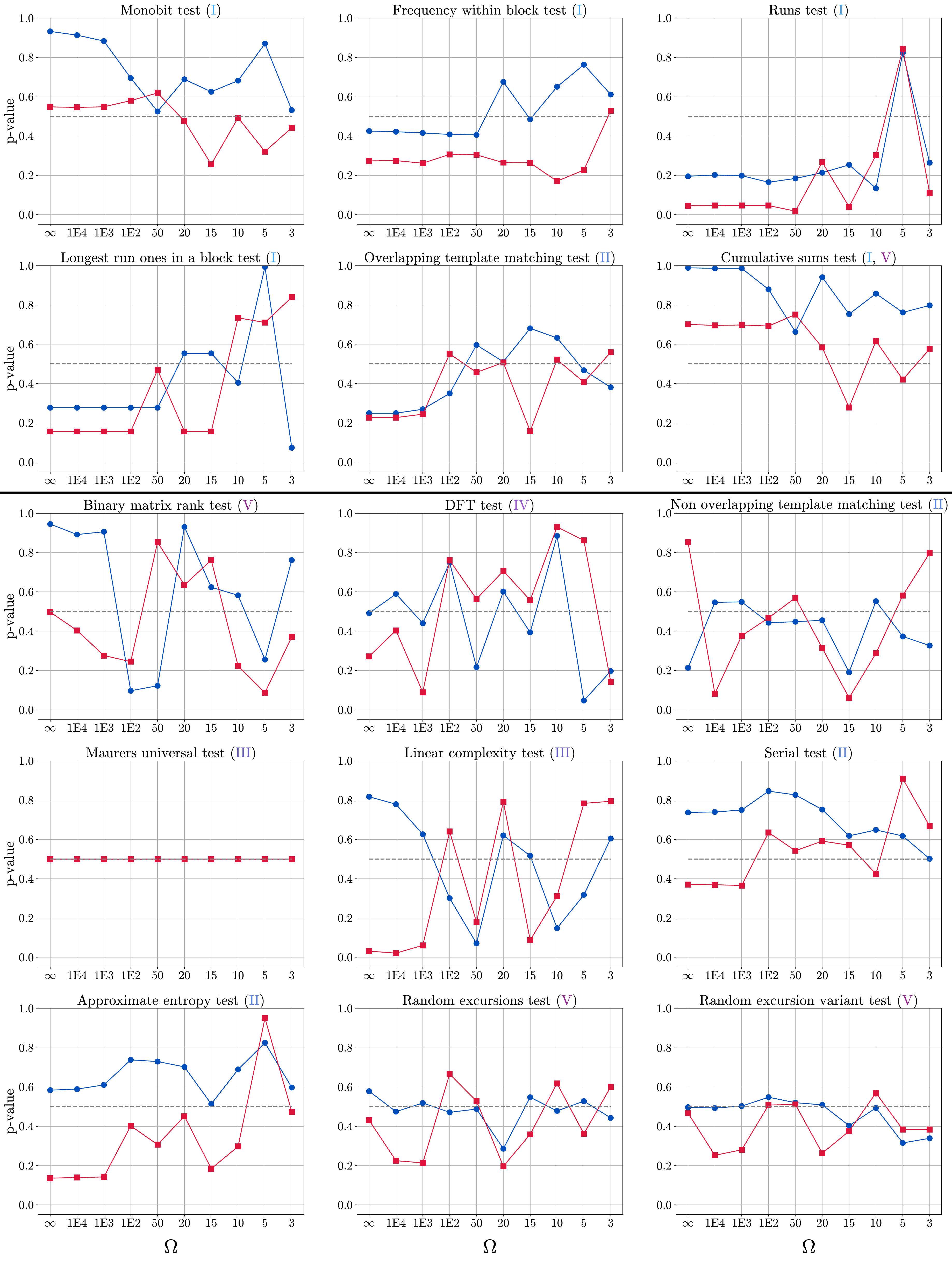}
    \caption{Randomness quality as a function of mixing frequency for: $\square$ (red) -- balanced coherent quasi-single photon source,  $\circ$ (blue) -- software-based ''Python'' source. The dashed line indicates the expected p-value =$0.5$. The tests represent following groups: (I) ''Balance'', (II) ''Template'', (III) ''Complexity'', (IV) ''Spectral'', and (V) ''Structural''. $\Omega$ represents the proportion of pseudorandom bits to genuine random bits in the mixed random number stream.}
    \label{Fig: mixing}
\end{figure*}

The technique involves adding with a certain frequency random bits from the high-entropy heralded RNG to bit sequences generated by the coherent quasi-single-photon and the software-based sources. We investigated the effect of varying the frequency of bit mixing on the quality of statistical tests performed on the obtained sequence. The results are shown in Fig.~\ref{Fig: mixing}.

\FloatBarrier

\bibliographystyle{quantum}
\newpage
\bibliography{bibliography}

%\enddocument
\end{document}